\documentclass[prb,aps,draft]{revtex4}

\begin{document}


\title{Field-Driven Hysteretic and Reversible Resistive 
Switch at the Ag-Pr$_{0.7}$Ca$_{0.3}$MnO$_{3}$ Interface}

\author{A.~Baikalov$^{1}$}
\author{Y.~Q.~Wang$^{1}$}
\author{B.~Shen$^{1}$}
\author{B.~Lorenz$^{1}$}
\author{S.~Tsui$^{1}$}
\author{Y.~Y.~Sun$^{1}$}
\author{Y.~Y.~Xue$^{1}$}
\author{C.~W.~Chu$^{1,2,3}$}
\affiliation{$^{1}$Department of Physics and Texas Center for Superconductivity, 
University 
of Houston, 202 Houston Science Center, Houston, Texas 77204-5002}
\affiliation{$^{2}$Lawrence Berkeley National Laboratory, 1 Cyclotron Road, 
Berkeley, California 94720}
\affiliation{$^{3}$Hong Kong University of Science and Technology, 
Hong Kong}

\date{\today}

\begin{abstract}
The hysteretic and reversible polarity-dependent resistive switch driven by 
electric pulses is studied in both 
Ag/Pr$_{0.7}$Ca$_{0.3}$MnO$_{3}$/YBa$_{2}$Cu$_{3}$O$_{7}$ 
sandwiches and single-layer Pr$_{0.7}$Ca$_{0.3}$MnO$_{3}$ strips. The data 
demonstrate that the switch takes place at the Ag-Pr$_{0.7}$Ca$_{0.3}$MnO$_{3}$ 
interface. A model, which 
describes the data well, is proposed. We further suggest that electrochemical 
diffusion is the cause for the switch.
\end{abstract}

\maketitle

Pr$_{0.7}$Ca$_{0.3}$MnO$_{3}$ (PCMO) has attracted extensive interest recently. 
Below 150~K, its free energies corresponding to the paramagnetic, the 
charge-ordered, and the ferromagnetic states differ only slightly. Therefore, a 
slight external disturbance, e.g. magnetic field, light, isotope mass, 
pressure, or electric field, may lead to a large resistivity ($\rho$) change, but 
only at low temperatures.\cite{asa}

Therefore, it is interesting to note the report of Liu \textit{et al.}\cite{liu} 
that the two-lead resistance, $R$, of a PCMO layer sandwiched between an Ag 
top-electrode and a YBa$_{2}$Cu$_{3}$O$_{7}$ (YBCO) or a Pt bottom-electrode can be 
drastically and repeatably alternated at room temperature by applying electric 
pulses with different polarities.\cite{note} This $R$-switch may thus offer 
potential device applications, e.g. nonvolatile memory. Similar 
$R$-changes in single-layer PCMO films with the four-lead configuration were also 
reported. The $R$-switch has therefore been attributed to bulk properties of 
PCMO, in terms of the alignment of the presumed ferromagnetic clusters by the 
electric field.\cite{liu} The interpretation, if confirmed, presents a major 
challenge to the physics of manganites and, possibly, to the basic law of parity 
conservation. The reported $R$-change of $\Delta R \ge 3000$~$\Omega$ across a 600~nm 
thick PCMO film represents a $\rho$-increase of 
$\Delta \rho \approx 10^{5}$~$\Omega$~cm, and suggests a novel state with a 
$\rho$(297~K) far greater than the 
$\rho$(297~K)~$<< 10^{1}$~$\Omega$~cm ever reported in PCMO. According to the 
commonly accepted polaron model, $\rho$(297~K) of manganites is controlled by the 
polaron mobility and should be ultimately restricted by the hopping barrier 
($10^{-1}$~eV~$\approx k_{B}T$ at 297~K) associated with the Jahn-Teller distortion, 
which is only a few eV.\cite{sal} The experimental $\rho$(297~K) is only $10^{-2}$ to 
$10^{0}$~$\Omega$~cm in (La$_{y}$Pr$_{1-y}$)$_{1-x}$Ca$_{x}$MnO$_{3}$ for 
$0.2 \le x \le 0.5$ and $0 \le y \le 0.7$,\cite{tom} and 
$\le 10^{4}$~$\Omega$~cm 
even in extreme cases, such as Nd$_{0.7}$Ba$_{0.3}$MnO$_{3}$ and 
LaMnO$_{3}$.\cite{coe} A $\rho$(297~K) of $10^{5}$~$\Omega$~cm or higher 
would 
suggest a new insulating state never observed before and challenge the polaron model 
commonly accepted. In a more general sense, this polarity-dependent $\rho$ 
in a uniform material reported, if proven, represents a violation of the law of 
parity conservation in the electromagnetic field. It may occur 
without parity violation only if the sample is asymmetric due to either an 
inhomogeneity in the thickness direction or poling by electric pulses 
(``training''); neither bears any obvious relation to the alignment model 
proposed.\cite{liu} The present study is motivated by our attempt to elucidate the 
mechanism responsible for, and the nature of, the $R$-switch. Our data demonstrate 
that the switch 
occurs at the Ag-PCMO interface, possibly via an electrochemical process. The 
observation not only resolves the parity-violation puzzle but also provides insight 
into optimization for possible applications.

PCMO films were synthesized by \textit{ac} sputtering at 760~$^{\circ}$C under a 
140~mtorr Ar:O$_{2}$ = 2:3 mixture atmosphere. YBCO films were synthesized by 
conventional pulsed laser deposition on LaAlO$_{3}$ (LAO) substrates. The structure 
was determined by X-ray diffraction (XRD) using a Rigaku DMAX-IIIB diffractometer. 
Both highly $c$-oriented and epitaxial PCMO films have been obtained and tested, with 
similar results. High voltage pulses were produced by a DEI PVX-4150 pulse 
generator. The resistivity was measured after applying the electric pulses by 
feeding a \textit{dc} current from a Keithley 2400 current source and reading the 
corresponding voltages using a HP34401A multimeter through a Keithley 705 scanner. 
Both the $\rho$(297~K)~$\approx$ 0.3~$\Omega$~cm and the $H$-induced 
metal-insulator transition around 120~K above 1~T observed before pulsing show that 
the PCMO films so prepared are typical of those previously examined.\cite{sin}

The Ag/(500~nm~PCMO)/(500~nm~YBCO) sandwich similar to that used by Liu 
\textit{et al.}\cite{liu} was first examined. Ag pads of 0.8 mm diam were deposited 
on the tops of both PCMO and YBCO (inset, Fig.~\ref{fig:fig1}). It should be noted 
that 
the two-lead $R$ varies greatly from pad to pad even before pulsing, in agreement 
with a previous report.\cite{liu} Additional 
three-lead measurements suggest that neither the YBCO layer nor the Ag-YBCO 
interface contributes significantly to $R$. Pulses of 30~V and 200~ns were applied 
between pads $A$ and $C$ (inset, Fig.~\ref{fig:fig1}). The pulse current density 
across 
both the interfaces and the PCMO layer was typically $\le 30$~A/cm$^{2}$, a range 
adopted to cover the threshold estimated from previous data.\cite{liu,note} A 
measurement current was applied after each pulse to determine the two terminal 
$R$'s. A possible thermoelectric effect on $R$ was eliminated by switching the 
polarity of the measuring current. The two-lead $R$ observed exhibits the same 
polarity-dependent switch previously reported.\cite{liu}

To determine whether the apparent parity-violation is associated with the electric 
poling, we simultaneously determined the voltages $V_{A-C}$ and $V_{B-C}$ across 
electrodes $A$--$C$ and $B$--$C$, respectively, with both the pulses and the 
measuring 
\textit{dc} current $I_{A-B} = 1$~$\mu$A passing through $A$--$B$. This configuration 
represents two serially connected sandwiches since the PCMO resistance between $A$ 
and 
$B$ is $10^{7}$ times higher than that perpendicular to the film. Both 
$V_{A-C}/I_{A-B}$ and $V_{B-C}/I_{A-B}$ are therefore expected to vary in the 
same fashion by pulsing if the pulse has generated a change in the bulk of the PCMO 
film, as the two sandwiches experience the same poling history. 
However, we found that $R_{A-C}$ increases while $R_{B-C}$ decreases 
(Fig.~\ref{fig:fig1}), depending on the polarity of the electrode. The apparent 
parity violation and the associated switches, therefore, cannot be attributed to the 
poling.

To explore the issue further, a single-layer 3 mm wide and 500 nm thick PCMO strip 
was measured in a multi-lead configuration with a 0.18 mm distance between the 
adjacent $0.32 \times 3$~mm$^{2}$ Ag pads (inset, Fig.~\ref{fig:fig2}a). A train 
of 20 pulses of 150 V and 200 ns with a given polarity was applied between 
electrodes 3 and 4 at a specific time. The estimated current densities through the 
two pads and through the PCMO strip are 40 and $10^{4}$~A/cm$^{2}$, respectively. 
Both are higher than the switching threshold $\approx$ 10--20 A/cm$^{2}$ estimated 
from both our data and the data of Ref.~\onlinecite{liu}.\cite{liu,note} After the 
electric pulses were applied, the transport elements $V_{l-k}/I_{m-n}$ were 
measured, where $V_{l-k}$ is the voltage between electrodes ($l$,$k$); and 
$I_{m-n} = 1$~$\mu$A is the measuring current through electrodes ($m$,$n$) with 
$l$, $k$, $m$, and $n$ = 1--6. Indeed, pulse-induced polarity-dependent switches were 
detected, as shown in Figs.~\ref{fig:fig2}a--b. The data demonstrate that the PCMO 
resistivity cannot be the dominant factor. First, the change of $V_{3-4}$/$I_{3-4}$ 
(i.e. the two-lead resistance across the pulse path) with the pulses is only 10\%, 
which limits the possible change of PCMO $\rho$(297~K) to below 
0.1~$\Omega$~cm (Fig.~\ref{fig:fig2}a). Such a change will only modify the 
$V_{B-C}/I_{A-B}$ in Fig.~\ref{fig:fig1} by $5×\times 10^{-4}$~$\Omega$, a million 
times lower than the 30~k$\Omega$ observed. Second, a similar switch was also 
observed in $V_{4-5}/I_{3-4}$, the contact resistance beneath pad~4 
(Fig.~\ref{fig:fig2}a and its inset). Therefore, contributions from the PCMO 
$\rho$(297~K), as the difference between the two-lead $R$ and the contact 
resistance, should only be $(V_{3-4}-V_{4-5})/I_{3-4} < 10$~$\Omega$ and within our 
experimental resolution.

However, a simple contact resistance model contradicts the observed switches in the 
conventional ``four-lead resistance,'' i.e. $V_{3-4}/I_{1-6}$, in addition to the 
interface resistances of $V_{2-3}$/$I_{1-6}$ and $V_{4-5}$/$I_{1-6}$. We attribute 
the contradiction to the non-zero size of the Ag pads: the large PCMO resistance 
may have forced the measuring current to detour through the Ag pads. A simplified 
resistive equivalent circuit for the Ag pads and the PCMO resistance underneath the 
pads is given in the inset of Fig.~\ref{fig:fig3}, where, for example, $R_{2A}$ and 
$R_{2B}$ are the effective interface resistances to PCMO on the left and right sides 
of pad~2, respectively; $R_{2C}$ is the the resistance for the PCMO section between 
these two effective contact points; and $R_{2D}$ is the PCMO resistance between 
pads~2 and 3. Changes of $R_{2A}/R_{2B}$, therefore, naturally lead to the change of 
$V_{3-4}/I_{1-6}$. A procedure was thus developed to simultaneously calculate all 
measured $V_{l-k}/I_{m-n}$ by assuming that only $R_{3B}$ and $R_{4A}$, i.e. the 
contact resistances along the pulse path, are affected by the pulses, and that all 
other contact resistances, which are not crucial to the data-fit, are equal and 
independent of the experimental sequence number $SN$ (inset, Fig.~\ref{fig:fig3}). 
The seven $SN$-independent parameters and the two $SN$-dependent parameters, 
therefore, can be uniquely deduced based on the 12 $SN$-dependent 
$V_{l-k}/I_{m-n}$ observed. The deduced $R_{3B}$ and $R_{4A}$ are shown in 
Fig.~\ref{fig:fig3}. The calculated $V_{3-4}$/$I_{3-4}$, $V_{4-5}/I_{3-4}$, 
$V_{3-4}/I_{1-6}$, and $V_{4-5}/I_{1-6}$, as a few samples of the 12 
$V_{l-k}/I_{m-n}$, are shown in Figs.~\ref{fig:fig2}a--b as solid lines. The 
agreement is good with only two variables of $R_{3B}$ and $R_{4A}$. Additional 
calculations further demonstrate that the additional contribution from PCMO 
$\rho$(297~K) should be less than a few percent of the resistance changes observed.

Very recently,\cite{tsu} we examined the pulse-induced resistive switch in 
several PCMO ceramic/single-crystal samples with either Ag films or Ag epoxy 
contacts as electrodes. Similar switches were observed. Both the two-lead resistance 
and the contact resistance are typically switched from $10^{2}$ to $10^{3}$~$\Omega$, 
similar to the data discussed above, although the measured four-lead resistance 
$\approx$~1--10~$\Omega$ is negligibly small and cannot account for the larger 
two-lead resistance switch observed. This is in perfect agreement with the 
proposition that the field-induced $R$-switch does not occur in the bulk of PCMO 
but at the interface between the electrode and the PCMO.

The electric field across the interface will be on the order of $10^{7}$ V/cm based 
on the voltage drops in the sandwiches and an assumed interface-thickness of 10~nm. 
This field is on about the same order as that expected for field-driven 
ion-diffusions. The relatively slow dynamics (Fig.~\ref{fig:fig1}), which can be 
improved by applying higher voltages, suggests electrochemistry, e.g oxygen 
diffusion or interfacial material change, as a route to the hysteretic and 
polarity-dependent $R$-switch observed.

In summary, the electric pulse-induced polarity-dependent hysteretic and reversible 
$R$-switch is observed in both the Ag/PCMO/YBCO sandwich and the single-layer PCMO 
configurations. Detailed studies show that the large $R$-change does not take place 
in the PCMO film but at the Ag/PCMO interface. A model based on interfaces is 
proposed to account successfully for the polarity-dependent $R$-switch. We further 
propose that a field-induced electrochemical surface reaction appears to be the 
driving mechanism for the switch.

\begin{acknowledgments}
We thank R.~L. Meng for the YBCO films. The work in Houston is supported in part by 
NSF Grant No. DMR-9804325, the T.~L.~L. Temple Foundation, the John J. and Rebecca 
Moores Endowment, and the State of Texas through the Texas Center for 
Superconductivity at the University of Houston; and at Lawrence Berkeley 
Laboratory by the Director, Office of Science, Office of Basic Energy Sciences, 
Division of Materials Sciences and Engineering of the U.S. Department of Energy 
under Contract No. DE-AC03-76SF00098.
\end{acknowledgments}

\break

\begin{figure}
\caption{\label{fig:fig1}$R_{A-C}$ ($\bullet$, $\times 1/30$) and $R_{B-C}$ 
($\circ$). $SN$ stands for the experimental sequence number. Individual $\pm 30$~V 
pulses are represented as bars at the bottom of the figure; two \textit{dc} 
$R$-measurements were carried out after each pulse.}
\caption{\label{fig:fig2}a) $V_{3-4}/I_{3-4}$ ($\bigtriangleup$) and 
$V_{4-5}/I_{3-4}$ ($\bigtriangledown$). Solid lines: calculated (see text). 
Inset: the electrode arrangement. 
b) $V_{3-4}/I_{1-6}$ ($\bigtriangleup$). Solid lines: calculated. The 
$\pm 150$~V pulses across electrodes 3 and 4 are shown as bars at the bottom. Each 
bar represents 20 pulses of 150~V and 200~ns.}
\caption{\label{fig:fig3}The model-calculated $R_{3B}$ (dashed line) and $R_{4A}$ 
(solid 
line). Inset: the proposed equivalent circuit.}
\end{figure}


\begin{thebibliography}{99}
\bibitem{note} Although the phenomena may actually be induced by electric fields, 
the local fields are difficult to calculate from the two-lead resistance. The 
overall $R$ includes the contributions from the Ag-PCMO interface, the Ag-YBCO 
interface, the bulk PCMO, and the bulk YBCO. The current, which should be the same 
in the four parts, is a better parameter. The threshold current density for both 
the interfaces and the PCMO is estimated to be 10~A/cm$^{2}$ from the data of 
Ref.~\onlinecite{liu}.
\end{thebibliography}

\end{document}